\documentclass[conference]{IEEEtran}
\IEEEoverridecommandlockouts
\usepackage{cite}
\usepackage{svg}
\usepackage{amsmath,amssymb,amsfonts}
\usepackage{algorithmic}
\usepackage{float}
\usepackage{svg-extract}
\usepackage{graphicx}
\usepackage{textcomp}
\usepackage{xcolor}
\usepackage{glossaries}
\def\BibTeX{{\rm B\kern-.05em{\sc i\kern-.025em b}\kern-.08em
    T\kern-.1667em\lower.7ex\hbox{E}\kern-.125emX}}
    
\newacronym{adhd}{ADHD}{Attention Deficit Hyperactivity Disorder}
\newacronym{id}{ID}{intellectual disabilities}

\begin{document}

%\title{Virtual Agent Tutors in Sheltered Workshops: Evaluating Feasibility for Attention Training in Individuals with Intellectual Disabilities}
\title{Virtual Agent Tutors in Sheltered Workshops: A Feasibility Study on Attention Training for Individuals with Intellectual Disabilities}

\author{\IEEEauthorblockN{Julian Leichert$^{1,2*}$}

\thanks{$^{1}$Medical Assistance Systems, Medical School OWL and $^{2}$Center for Cognitive Interaction Technology, CITEC from Bielefeld University, Germany. $^{*}$Corresponding Author: Julian Leichert \tt\small jleichert@uni-bielefeld.de.}

\and
\IEEEauthorblockN{Monique Koke$^{1,2*}$}

\and
\IEEEauthorblockN{Britta Wrede$^{1,2*}$}

\and
\IEEEauthorblockN{Birte Richter$^{1,2*}$}

}

\maketitle

\begin{abstract}
In this work, we evaluate the feasibility of socially assistive virtual agent-based cognitive training for people with \gls{id} in a sheltered workshop. The RoboCamp system, originally developed for children with \gls{adhd}, is adapted based on the results of a pilot study in which we identified barriers and collected feedback from workshop staff. In a subsequent study, we investigate the aspects of usability, technical reliability, attention training capabilities and novelty effect in the feasibility of integrating the \textit{RoboCamp} system.

\end{abstract}


\section{Introduction}
% people with id
People with \gls{id} are an important part of the general population, with a prevalence rate of 1\% worldwide, 0.5 to 1 million people in Germany alone \cite{sappok_medical_2019}. In Germany, there are 310,000 people with disabilities employed in sheltered workshops. Around 75.03\% of these employees have an intellectual disability \cite{bagwfbm2023}.
One of the daily challenges they face is the allocation of attentional resources \cite{garcia-pintor_attentional_2024}, which can lead to difficulties in everyday activities and workplace tasks, many of which require sustained focus. 
% ADHD and people with ID
According to \cite{spaniol_meta-analysis_2022}, people with \gls{id} perform lower on executive function tasks, including visuospatial attention tasks.
Furthermore, \gls{adhd} occurs more frequently in the population of people with \gls{id} than in the general population \cite{baker2010mental}.
Even though the prevalence is higher, the disorder is often underdiagnosed and/or misdiagnosed \cite{miller2020clinical}.
Training attention maintenance could enhance the required abilities for a transfer to the job market, ultimately improving employability and independence.

Social robots and virtual agents have gained attention for their potential in cognitive training and virtual tutoring in multiple settings. They offer promising solutions by alleviating the burden on caregivers and providing engaging, autonomous opportunities for the development of cognitive skills in a structured setting. \cite{westra_evaluating_nodate}
Despite the growing body of research on assistive technologies for individuals with \gls{id}, there remains relatively limited investigation into the feasibility of cognitive training programs for this population, particularly those utilizing socially assistive virtual agents. This paper addresses the following research question
(RQ): \emph{Is the integration of a virtual agent-based assisted attention training in a sheltered workshop feasible?}

To explore this, we focus on improving and adapting an existing attention training program, \emph{RoboCamp}, which was originally developed for children with \gls{adhd}. Through a series of iterative improvements, the program is tailored to meet the specific needs of individuals with \gls{id} in sheltered workshops. 
We evaluate the feasibility of the system across multiple dimensions, including technical reliability, usability, the novelty effect, and the system’s ability to train various aspects of attention within a heterogeneous group of individuals, many of whom have comorbidities. 
The goal of this work is to assess the potential of the system for real-world deployment in sheltered workshop settings.
%We examine the integration of the system regarding the aspects of technical reliability, usability, the novelty effect, and the system's ability to train different aspects of attention in a heterogeneous group of individuals with comorbidities, we aimed to assess the feasibility of the system in real-world conditions. 

 \begin{figure}[H]
        \centering
        \includegraphics[width=0.46\textwidth]{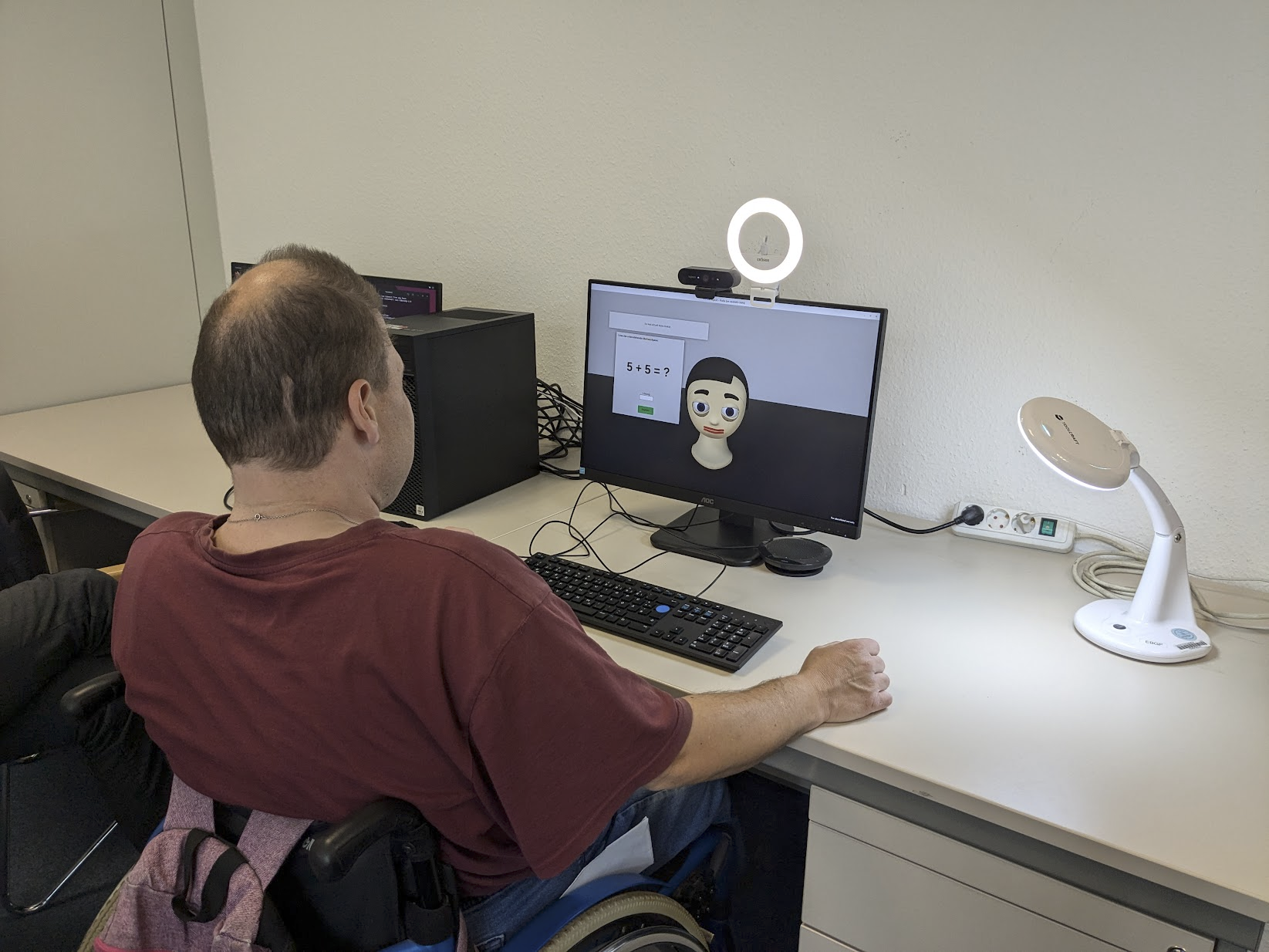}
        \caption{Employee interacting with the \textit{RoboCamp}}
     \label{fig:robocamp}  
\end{figure}

\section{Background and Related Work}
To enable effective training of cognitive functions, in this case attention, it is essential to understand the underlying concepts of attention itself. Existing approaches of agents assisting in cognitive training and their impact on users, especially the application in the group of people with intellectual disabilities (ID), are examined. Furthermore,  this section introduces the \textit{RoboCamp}, how gaze relates to attention training, how long term interactions impact performance, and concludes with the gaps we found in current research in this domain.

\subsection{Attention}
% different types of attention
According to \cite{posner_attention_1990}, attention can be divided into three networks associated with neuroanatomical regions: orienting, alerting, and target detection, also referred to as executive control. Being vigilant for incoming stimuli is the responsibility of the alerting network, whereas the executive network manages the targeting of a specific stimulus. The orienting network's focus is to select a modality or location to prioritize sensory input.\\
The mechanisms of orienting our attention can be divided into overt and covert attention. Overt attention refers to the visual focus on a stimulus, whereas covert attention involves directing attention to something that is not in the visual focus—for example, listening to a conversation while fixating on another stimulus \cite{gazzaniga2019cognitive}.

Beyond neuroanatomical models, clinical frameworks provide a functional perspective on attention, such as the widely used model by \cite{sohlberg1987effectiveness}, which categorizes attention into five distinct types:

\begin{itemize}
\item Sustained Attention: Maintaining a consistent behavioral response during continuous or repetitive activity.
\item Selective Attention: Focusing on a stimulus while ignoring other stimuli.
\item Focused Attention: Responding directly to a stimulus.
\item Alternating Attention: Mentally switching between tasks with different cognitive requirements.
\item Divided Attention: The ability to respond to multiple stimuli at once.
\end{itemize}

% Erwähnen von anderen hierarchischen Modellen

In contrast, the model used by \cite{sturm_leitlinie_2009} merges selective and focused attention into a single category, incorporates alertness, vigilance (sustained attention), and spatial attention, and unifies divided and alternating attention under attentional flexibility.

\subsection{Assisted Cognitive Training}
% cognitive training beinhaltet attention training 

Cognitive training involves various domains, including memory, reasoning and attention. 
The use of socially assistive agents is a promising novel approach to reduce burden on caregivers and facilitate engagement in users.

% neurotypical individuals and cognitive training
In neurotypical individuals, it has been shown that positive feedback given by social robots impacts task performance \cite{shiomi2020two}. As the acquisition of a robot comes with a high cost, virtual agents demonstrate a low-cost alternative as assistive technology in cognitive training. 
As shown by \cite{brachten_ability_2020}, they can decrease cognitive load in a given task.
\cite{zsoldos_value_2024} found that a virtual assistant was perceived as useful during computerized cognitive training (CCT) in a Wizard of Oz setup. 

% cognitive training for people with ID
However, there is relatively little research on assisted cognitive training for people with \gls{id}.
\cite{yuan_systematic_2021} conducted a review on the applications of social robotics in rehabilitation for cognitive training. They found that most studies were conducted in controlled lab settings and that many publications focus on \gls{adhd} and ASD. 
%There is only a limited number of works on assisted cognitive training for individuals with \gls{id}.
%Since the goal of RoboCamp is to train attentional capabilities, it is essential to define the concept of attention in this context.
% assistive systems for attention with people with ID
\cite{pashapoor_effectiveness_2018} found that cognitive computer games significantly improved the attention span of children with \gls{id} compared to a control group without training.

In their research, \cite{shukla_robot_2019} identified key aspects of robotic interventions for individuals with \gls{id} through expert interviews and conducted a case study with the NAO robot. In this study, NAO performed different types of interactions derived from expert interviews. Questionnaires showed that the interventions had positive effects on engagement of the users.

In a study in which an assistive robot was used to train episodic memory of children with \gls{id}, a positive effect on episodic memory abilities was demonstrated \cite{perez_effects_2021}. \cite{arora_employing_2022} presented a neurorehabilitation system composed of adaptive training exercises supported by a socially interactive agent. In a consecutive pilot study, they qualitatively evaluated the feasibility of the system and reported that participants adapted quickly to the system and were not distracted by the virtual agent. Statements of participants indicate a motivating factor \cite{garcia-pintor_attentional_2024}.
The cognitively assistive robot (CARMEN) was developed by \cite{bouzida_carmen_2024} in collaboration with clinicians and individuals with mild cognitive impairment (MCI) to autonomously deliver compensatory cognitive training interventions at home. In  a feasibility study, they found that participants liked CARMEN. The usability of the system was ranked average on a SUS questionnaire answered by participants and staff.

\cite{mois_understanding_2020} examined whether cognitive training led by a socially assistive robot is feasible for older adults with mild cognitive impairment. In their study, the Musical Assistive Robot Instructor (MARI) led piano lessons over four weeks and found that multiple cognitive functions of the participants improved. The participants perceived the robot as competent, moderately warm, and experienced minimal discomfort.

\textbf{Attention training:}
% Robocamp
The \textit{RoboCamp} therapy system was originally designed to help improve the concentration of children with \gls{adhd} through motivational attention training using a virtual robotic agent \cite{richter2024vaco}.
Its feedback system is based on the response cost token, developed by \cite{adhs-summercamp_2009}.

% gaze, eye tracking
While the tracking of visual attention is receiving increasing attention in research on \gls{adhd} in people without \gls{id} \cite{lee2023use,lev2022eye}, its specific application in people with \gls{id} remains unexplored. 
\cite{cavadini2022eye} demonstrated good feasibility
of the eye-tracking paradigm in this group of individuals, regardless of the presentation or content of the tasks. 
\cite{skvaznikov2019using} showed that adding information sources, such as eye-tracking data or emotion recognition, can increase the reliability of test results in people with \gls{id}.

\subsection{Long term interaction}
% long term interaction, novelty effect
However, while assistance systems in the real world have to work and be efficient over long periods of time, studies general focus on single encounters only. When investigating long-term repeated interactions, it is often the case that a decrease in interest, and thereby in performance can be observed over time. This is called the ``novelty effect'' \cite{fryer2017stimulating}. More specifically, in their study, \cite{fryer2017stimulating} showed that interest in the task decreased already after the 2nd interaction when interacting with a chatbot as compared to a human interaction partner. 

% Usability 
\subsection{Usability}
%usability: key aspects, evaluation
Not only the "novelty effect" \cite{fryer2017stimulating} influences the feasibility of integrating the attention trainer as a tool within the sheltered workshop. The system here has to be easy to learn, easy to use and easy to remember. Usability refers to the idea, that products should be designed in a way, that users see the benefit of the product (e.g. by achieving goals), know how to use it and enjoy using it. \cite{bastardo2021methodological} As \cite{bastardo2021methodological} state, the evaluation of usability does not follow a standardized procedure, which leads to problems in the quality of the results and therefore, their transferability. This is why, \cite{bastardo2021methodological} ask for more detailed descriptions on how usability is evaluated within studies to open the possibility of identifying aspects of the interaction that were not taken into account as relevant prior to the analysis.

\subsection{Research Gap}
% research gap
There is a gap in research on virtual agent-assisted cognitive training for people with \gls{id}, as much of the research in this domain focuses on ASD, elderly individuals with mild cognitive impairments, or children with neurodevelopmental problems. In the context of sheltered workshops, no research could be found on assisted cognitive training. It is this gap in research that we aim to address in this work.

\section{Pre-Study}
\begin{figure}[H]
        \centering
        \includegraphics[width=0.5\textwidth]{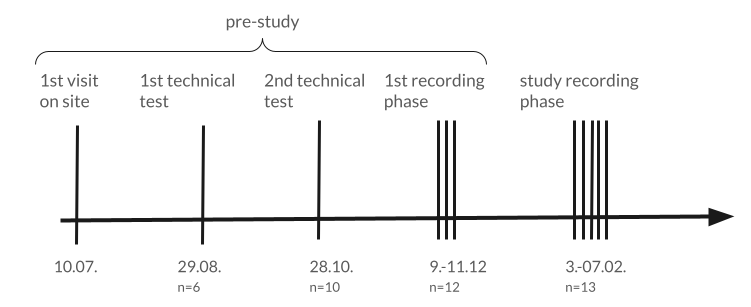}
        \caption{The five visit phases in the sheltered workshop in 2024 and 2025.}
     \label{fig:visits}  
\end{figure}
To identify potential barriers and necessary adaptations to transition from the previously targeted \gls{adhd} group to individuals with \gls{id}, we conducted a pre-study with employees and staff of the sheltered workshop. \textit{Employees} refer to workshop participants with \gls{id}, while \textit{staff}includes caregivers/supervisors employed at the workshop. 
The system was tested and refined through repeated visits (see \ref{fig:visits}), incorporating continuous feedback from the users. 
The first visits focused on investigating the technical requirements of the site and identifying potential locations for the system installation.
The goal of the second phase (visit phases 2 and 3) was to test the system's key functionalities, including lighting conditions and attention detection, while ensuring usability through staff interaction.
The final visits (visit phases 4 and 5) focused on recording interactions with people with \gls{id} and gathering feedback to further improve the system.

During the pre-study (visit phases 1-4), several observations were made regarding the system's technical performance and the usability. It was identified that lighting conditions needed improvement, as backlighting and inconsistent illumination affected the system's ability to accurately detect user interactions.
Additionally, participants had difficulties locating the correct keyboard button to submit their solutions. Staff pointed out that overly strong negative feedback could induce frustration. One participant was unable to correctly submit solutions, as the degree of physical disability did not allow entering numbers and pressing the enter key; only pressing the enter key was possible.

Another possible cause of frustration was identified: the continued submission of incorrect solutions could irritate employees.

\section{Study}
To address our research question and evaluate the feasibility of the system for individuals with \gls{id}, we conducted a study in a sheltered workshop. Incorporating insights from the pre-study, we adapted the system to mitigate the identified issues. The following sections outline the methodology, participant characteristics, system setup, adaptations made, and study procedures.

\subsection{Participants}\label{AA}
In total, 13 participants with different levels of intellectual disability participated in the study, some of the participants also had a physical disability. 
One of the participants who took part in the pre-study did not participate in the study due to the degree of physical disability. All participants were employees of the sheltered workshop. No participant was excluded based on age, sex, or disability to avoid rendering them vulnerable or victimizing them by assigning them to a particular group. 

\subsection{Setup}
The study was carried out using an Ubuntu-based PC equipped with two monitors, two webcams, a ring light, a speaker, a mouse, and a keyboard, as seen in Fig. \ref{fig:robocamp}. The primary monitor was used for participant interaction, displaying the tasks, points, and the agent Flobi. The secondary monitor was designated for system monitoring and placed in a way that was not directly visible to the participant to ensure that it did not interfere with their experience.

Two webcams were placed above and below the
monitor to capture participants head position and gaze direction for analysis of engagement/attention. The one on top was used for the live detection of attention, the bottom
one for later analysis on optimal webcam position. A ring
light was used to improve lighting conditions. The mouse and keyboard provided standard input
methods for the participant’s interactions with the system.
The keyboard enter key was marked with a blue dot to
enable participants to identify the required key for entering
the solution.

\subsection{Attention Detection}\label{attention detection}

To detect the direction of gaze, the OpenFace library \cite{baltrusaitis2018openface} is used to track eye movements. A threshold of +/-0.22  rad is applied in the x direction and -0.22  rad in the y direction. Positive values in the y direction are disregarded to prevent false inattention triggers when participants look at the keyboard. When the gaze direction succeeds the x or y thresholds for more than 30ms, the participant is considered inattentive by the system.
To enable synchronicity between OpenFace gaze detection and user input, both data streams were published as ROS topics and recorded in a rosbag.

\subsection{Interaction}\label{interaction}

The session starts with an explanation by Flobi of the reward system and how to interact with the system, as where the points are displayed, and which key to press. If the person starts early, it is rewarded with a point, if attentive as detected by the system, it is rewarded with points in intervals determined by the length of the session. The verbal feedback consists of statements such as, "You're doing well, another point for you", "Great you're attentive. Keep going." . When the person looks away is regarded as inattentive, as elaborated in \ref{attention detection}, Flobi intervenes with feedback to recapture the attention. These Feedbacks include statements such as "Great, you are concentrating again", and "Great, (participant's name), let's move on". 
During the session, when no feedback is deployed, Flobi looks at the tasks in specified time intervals to facilitate joint attention.
At the end of the session, Flobi thanks the person for participating and says goodbye.

\subsection{Measurements}
To assess the reliability of the system in terms of detecting lapses of attention, the impact on task performance and the ability to maintain attention, the following measurements were made.
The duration of attention/inattention, as detected by the system, how long participants take to redirect their attention to the task. The time and number tries needed to correctly solve the presented math tasks. The number of tries is restricted to two tries, to not irritate the participants with reoccurring wrong answers. The gaze angles computed by OpenFace were recorded, alongside the webcam recordings of one webcam placed on top of the screen, and one below the screen for later evaluation of the optimal webcam position for detecting the face and gaze.

\subsection{Conduct}
The sessions with the participants were conducted over a period of 5 days, each participant attended one session of 20 minutes per day, on the first day a baseline recording was made without verbal feedback from Flobi.
Due to different working times, and sick leaves, not all participants completed all planned 5 sessions.
In the first session, Flobi only introduced the exercise and remained silent for the rest of the study. Flobi's background behavior, i.e. looking at the exercise and blinking, occurred during both types of session. In all subsequent sessions, Flobi acted as described in \ref{interaction}. In the last two days, one of the staff members executed the recordings without supervision. Including, starting/stopping the training sessions and selecting a corresponding configuration. All measurements and recordings were started programmatically at the beginning of a session.

\section{Results}
In the following, the results of the conducted study are laid out. They include feedback by participants and staff, the performance of the attention detection, and the results of the training itself. We excluded participants who only completed the first training session, where no feedback of Flobi was deployed. One participant was left out of the analysis, as the recordings stopped early due to a wrong time configuration.

\subsection{Feedback by Participants and Staff}

\begin{figure}[H]
        \centering
        \def\svgwidth{0.4\textwidth}
        \includesvg{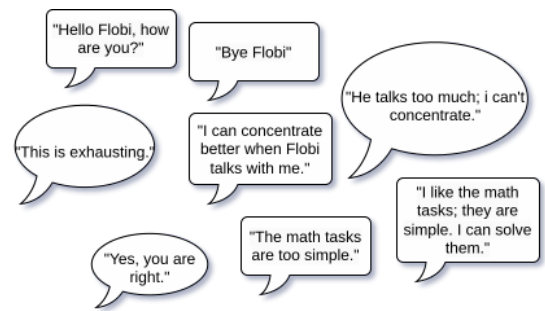}
        \svgsetup{inkscapelatex=false}
        \caption{Verbal statements by the participants during and after the sessions.}
     \label{fig:bubbles}  
\end{figure}

Participants and caregivers made some verbal statements during the study, which are depicted in Fig. \ref{fig:bubbles}.
An employee caregiver mentioned that one of the participants was more attentive after the interventions. Whilst working on task in the workshop, the participant said he has to be attentive, just like in the \textit{RoboCamp}.

Three of the participants were very eager to continue the \textit{RoboCamp} training as the study was propagated by the staff at the workshop and word spread around the people working there. Some participants tried the \textit{RoboCamp} due to sheer interest in the new system. 
Noteworthy is that board members of the sheltered workshop and the provider are open to further studies and see potential in other areas, such as job allocation and onboarding assessments of cognitive capabilities.

Alongside verbal feedback, which was provided during the sessions, hinting at the users' perception of the system,  we analyzed one participant-Flobi-interaction throughout all sessions to gain first insights into usability of the system, as well as evaluating formal aspect of the \textit{RoboCamp} training with respect to usability.
 On a formal level, the \textit{RoboCamp} training consists of three phases (see: Subsection IV.D for details): 1) introduction; 2) task performance and 3) end of training. This process was repeated in the same way every session. Therefore, the possibility of familiarization processes to occur is given. With respect to the usability, this indicates that the system should be predictable for users after a few sessions. Increasing familiarity is likely to go along with an increased feeling of safety while interacting with the system, which would also positively influence usability. When the system becomes as predictable for the user as the structure indicates, users might increasingly focus on the task at hand, rather than the interaction with the virtual agent. If these effects occur, they should be observable when analyzing the user's behavior while training.

\textit{Behavior during Session 1: }
During the interaction phase, the participant was highly engaged with the virtual agent, showing social signals like nodding or giving a thumbs up as a reaction to Flobi's explanation of the following procedure. This non-verbal behavior was accompanied by long gaze fixation to the virtual agent and facial expressions that indicate a high level of concentration throughout the interaction. Both aspects taken into account, first exposure to the system seems to impose that a high demand of focused attention is needed to gather all relevant information for the training session. In cases of severe intellectual disability, this might have negative effects on usability. Nevertheless, verbal input, including explanations on how to use the system and the goal of the interaction, given by the virtual agent appears to be easy to understand for the participant, as they directly start to uptake the active role of providing answers and attention during task performance. 
At first, the participant showed signs of irritation when receiving (erroneous) feedback from Flobi while looking down at the keyboard for a prolonged period of time to enter the solution. This happened several times until the participant changed their behavior: When entering the solution, they decreased their gaze towards the keyboard. Instead, they enforced their gaze towards the screen. This adaptation of behavior exhibits first learning mechanisms in the user and provides insight into the transparency of the system from the user's perspective: The participant was able to adapt their behavior, probably because they were able to notice the underlying feedback scheme within a few feedback loops (Participant looks away for a prolonged time span. Flobi provides feedback on lost points. \textit{vs.} Participant looks at Flobi. Flobi provides points.). The relevant interaction basics and contextual rules seem to be transparent enough to be noted within the interaction, and when identified, easy to learn. However, this trial-and-error-approach might create problems for some users who cannot identify those rules without direct explanation. At the end, the participant missed Flobi's ending words, which reflects back to attention rather than usability. 

\textit{Behavior during Session 3:} 
Different from session 1, the participant does not seem to be highly engaged in this introduction phase. They present a high level of body movement (like turning the upper body from left to right) and often look somewhere else during Flobi's explanation. Both factors indicate a low focus on the explanation of the system. Besides, the participant exhibits gestures of unease and discomfort, like forcefully pushing air through their lips while touching their neck. At first glance, this behavior can be interpreted as disengagement and boredom, hinting at negative feelings towards the interaction with the agent. However, the transition from introduction to task performance paints a different picture: Completely changing the level of engagement from fairly not existent to highly engaged, the participant starts to solve the task during the second phase. Their gaze is completely fixated on the screen, fastly typing in the solutions. Taking this engagement switch into account, the disengagement from the beginning hints at aspects of usability: The participant already knows what Flobi is going to say and what to do. This shows, that the information given in the explanation phase is not only easy to learn, but also easy to remember. So much so, that it almost bores regular users. Interestingly, the participant was able to divide attention between the ongoing exercise and the demand of typing in the solution whilst answering to a question that was asked by a staff member throughout training. One could argue, that this reflects on usability: If users are able to multitask while interacting with the system and still are able to talk to someone at the same time, the demands posed within the interaction are well manageable for users. Towards the end, the participant seems to expect the end of the training, repeatedly looking at their watch, leaning backwards, showing an excited facial expression while pushing up their fists to the air. When realizing, that the training is still ongoing, his whole body "sinks in". Their face shows disappointment, when continuing the task. At the end, the participant seems happy to have finished the training and leaves immediately. It seems, training sessions are perceived as long. However, the participant depicts a "feeling of predictability of the system" at this point. 

From analyzing session 1 and 3 we can summarize for the usability of the system with respect to this participant: 
\begin{itemize}
    \item The system seems to be easy to use.
    \item The "\textit{how?}" and "\textit{what?}" of the interaction seem easy to understand and learn.
    \item The system seems to be easy to remember.
    \item The training is predicable, therefore might give a feeling of safety in use.
    \item The system can be used independently. 
\end{itemize}

\subsection{Performance of Attention Detection}
To ensure a positive interaction with the virtual agent Flobi, it is important to detect attention reliably. OpenFace tracked facial features well for most participants, but some had problems due to head positioning. Participants who often had their heads tilted towards the keyboard blocked the view of their eyes and facial features for the top-mounted camera. This resulted in incorrect face recognition. One participant in particular kept looking down, which caused frequent instances of detected attention loss because OpenFace couldn't detect his face.

Looking at the average gaze angle for each participant, it is clear that the center of attention does not coincide with the midpoint of the outer limits of the x-axis threshold. 

 \begin{figure}[H]
\begin{minipage}{.5\textwidth}
    
    \centering
    \def\svgwidth{0.9\textwidth}
    \svgsetup{inkscapelatex=false}

    \includesvg{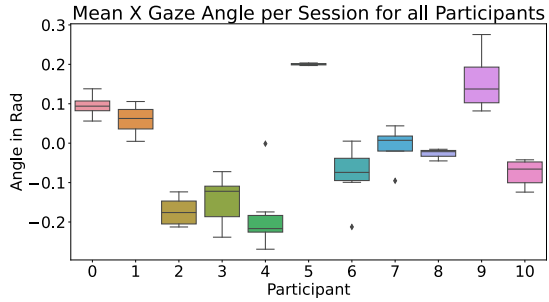}
    \caption{Mean x gaze angle over sessions for each participant}
    \label{fig:mean_x}

\end{minipage}
\end{figure}

\begin{figure}[H]
\begin{minipage}{.5\textwidth}

    \centering
    \def\svgwidth{0.9\textwidth}
    \svgsetup{inkscapelatex=false}

    \includesvg{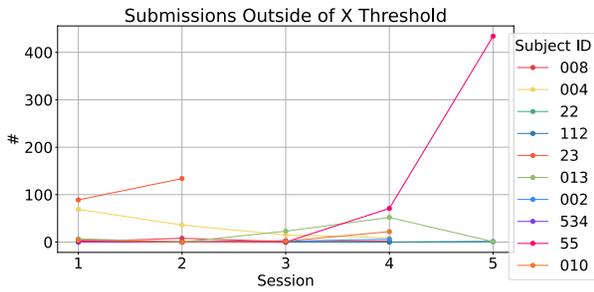}
    \caption{Number of Submissions per participant outside of gaze angle x threshold}
    \label{fig:outside_x}

\end{minipage}%
\end{figure}

Only 3 participants had an average x-gaze angle close to 0, as shown in Fig. \ref{fig:mean_x}. In the y-direction, only one participant has an outlier outside the threshold.
When examining the timestamps of submitted correct solutions and the direction in which the person is looking, it is possible to detect occurrences, as depicted in Fig. \ref{fig:outside_x} where the x gaze direction succeeded the selected thresholds while the participant is submitting a solution.

To fully understand these instances, further analysis of the video stream captured by the bottom-mounted webcam is required.

\subsection{Performance of the Participants}

\begin{figure}[H]
\begin{minipage}{.5\textwidth}
    
    \centering
    \def\svgwidth{0.9\textwidth}
    \svgsetup{inkscapelatex=false}

    \includesvg{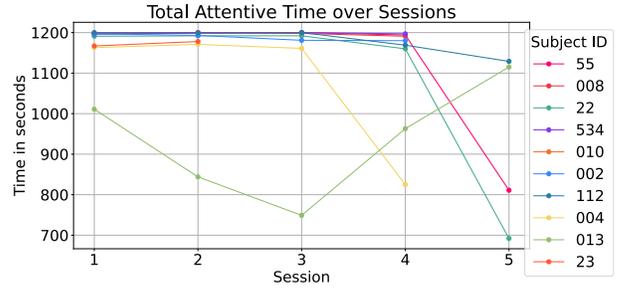}
    \caption{Attentiveness of participants across sessions}
    \label{fig:attentive}
\end{minipage}
\end{figure}

\begin{figure}[H]
\begin{minipage}{.5\textwidth}

    \centering
    \def\svgwidth{0.9\textwidth}
    \svgsetup{inkscapelatex=false}

    \includesvg{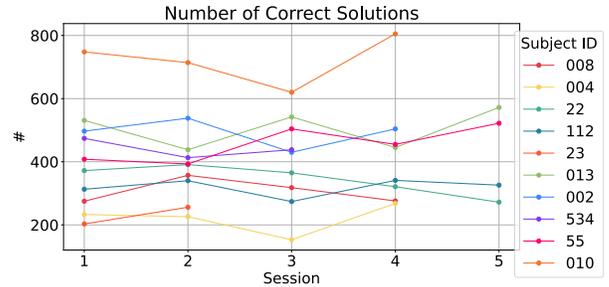}
    \caption{Number of Correct Solutions for Participants per Session}
    \label{fig:correct_sol}
\end{minipage}%
\end{figure}

The total duration of attentive time remains high for most participants over the course of four sessions, as seen in Fig. \ref{fig:attentive}. A slight downward trend is noticeable in the first four sessions for three participants. Participant 004 experienced a substantial decrease in attention duration from session 3 to session 4. For one participant (013), a clear downward trend is visible until session 3, followed by an upward trend in sessions 4 and 5.
In the final session, there is a drop in attention duration for all participants on that day, except one (013). For the four participants who took part in all 5 sessions, the mean difference between the first and last session is 180.4.
Whereas, the mean difference in number of correct solutions between the first and last session of those four participants is -17.0.
The task performance of the participants over sessions, measured in number of correct solutions is seen in Fig. \ref{fig:correct_sol}.
One participant showed a clear downward trend after the second session. Another showed consistently much higher performance in the task than other participants. For five participants, more correct solutions were recorded in the second session than in the first no-feedback session. For most participants, the performance fluctuated over five sessions.

%\begin{figure}[H]
%\begin{minipage}{.5\textwidth}

%    \centering
%    \def\svgwidth{0.9\textwidth}
%    \svgsetup{inkscapelatex=false}

%    \includesvg{figures/mean_task.svg}

%\end{minipage}%
%\end{figure}
\section{Discussion}
The following section reflects on the main findings of our study in relation to RQ1, focusing on the key aspects of usability, the novelty effect, technical reliability, and the training of attention.

\subsection{Usability}
One aspect we looked at was whether the system could be used by the heterogeneous group of people with ID in this workshop.
As this group includes people with co-morbidities such as severe physical disability and intellectual disability, accessibility is a critical aspect of this system.
Only one person was not able to participate in the (pre-)study because the input modality of the keyboard was not feasible for the degree of physical disability. In addition, the position of her head did not allow correct detection of attention, which made the feedback from Flobi useless and disabling in this case.
For those who were not severely physically disabled, it was possible to use the system without barriers. Staff were able to select the correct configuration for a participant, start, and stop sessions after a short introduction to the system, which speaks for usability on the caregiver side.
%-Feedback
%-heterogenität der Zielgruppe: qualitativer Ansatz: Kann man via Analyse der %Auseinandersetzung mit dem System auf dessen Nutzbarkeit zurückschließen? 

\textit{Qualitative approach to evaluating the usability of the system within our context}

As before mentioned, the implementation of the system within the sheltered workshop was conducted with the help of live online feedback. Nevertheless, the target group is highly heterogeneous. This means, that some methods like questionnaires or interviews on assessing the usability of a system might work on some participants whereas others could not participate in this method solely based on the severity of their intellectual disability. Therefore, we used a qualitative approach to see, if it was possible to evaluate the usability of the system through participants' interaction with the system itself. Our first results seem to indicate that further analysis could provide an empirical base for quantitative evaluation of the usability of the \textit{RoboCamp} training within our context. 

\subsection{Technical Reliability}
Attention detection functioned accurately in most sessions; however, certain factors can pose challenges to its effectiveness. Occlusion caused by tilted head positions can make it difficult to correctly detect a person’s face, potentially leading to misinterpretations of attention levels and resulting in erroneous feedback. To mitigate such errors, multiple modalities could be integrated to improve the system’s confidence in its assessments. Additionally, variations in sitting positions can lead to different gaze directions, which the system might misclassify as inattention, increasing the likelihood of false positives.

\subsection{Novelty Effect}
The results shown in Fig. \ref{fig:attentive} indicate for most participants that the attention remains high for most of the sessions. Only for the last session, a clear drop can be observed.
This seems to be an indicator that there is no novelty effect after the first or second session, as often seen in other studies (e.g. \cite{fryer2017stimulating}). The drop in attention at the last session might, however, be a somewhat delayed novelty effect. This delay might be a consequence of the special population with different levels of intellectual disabilities, which can leave to a slowed effect. Another explanation for this drop might be the fact, that the 5th session took place on a Friday. In a survey on the effect of work day on output and performance, \cite{bryson2007there} report more studies that find a performance drop towards the end of the week as opposed to studies finding an increase. Thus, the drop in attention time might be an effect of fatigue accumulating over the week and being released at the last day of the week.
Most interestingly, in contrast to this stark drop in attentive time, we find no such effect in the number of correct solutions (cf. Fig. \ref{fig:correct_sol}). While we do see individual differences between participants, there are only moderate variations in most participants over the different sessions. While it remains unclear why there is no correlation between attentive time and number of correct solutions, this might be an indicator for the absence of a novelty effect. 
Further research should address this issue by taking into account even longer time spans and more repetitions.

\subsection{Training Aspects of Attention}
There were not enough sessions to make a clear statement about attention trends. The math tasks used in the study could be linked to both selective and sustained attention, requiring users to maintain focus over time while also suppressing potential distractions, such as chatter of colleagues appearing at the open door. Although these tasks were suitable for all participants, variations in performance highlight the need for adaptability, allowing tasks to be adjusted in difficulty to suit the user. As almost all participants showed high total durations of overt attention, increasing the duration of the session could provide insights into sustained attention capabilities.

\section{Summary and Future Work}

In this contribution, we successfully implemented virtual agent-assisted cognitive training for people with \gls{id} in a sheltered workshop environment. 
%We addressed gaps in research, such as the lack of in-the-wild experiments with virtual agents in sheltered workshops and agent-assisted cognitive training for people with \gls{id}.
Our work contributes to addressing critical gaps in the literature, such as the scarcity of in-the-wild experiments involving virtual agents in sheltered workshops and the limited application of agent-assisted cognitive training for individuals with \gls{id}.

Regarding the research question — \emph{Is the integration of a virtual agent-based assisted attention training in a sheltered workshop feasible?} — the findings demonstrate that the system is indeed feasible. 
Feedback from both participants (employees) and staff, coupled with their continued interest, and successful use of the system over a one-week period without expert supervision, indicates that the system can be effectively integrated into sheltered workshop settings.
%their ability to use the system successfully over the span of a week, even without expert supervision. 
While initial findings suggest that the training may be beneficial for some participants, the duration of the study (one week) was insufficient to fully assess its impact on attentional capabilities. A longer-term study is required to investigate whether sustained training can lead to measurable improvements in attention for individuals with \gls{id}.
%We found initial indications that training could be effective for some participants, but a week was not sufficient to accurately measure attentional capabilities. To investigate whether long-term training will affect the participants' attention capabilities, a long-term study is necessary.
%To overcome technical issues, such as malfunctions in face detection, different face recognition models can be examined for their performance. The impact of head positioning could be addressed by introducing a calibration of thresholds before the session or an adaptive approach based on user input.
Several technical issues were encountered during the study, particularly with face detection malfunctions. To address these, further exploration of alternative face recognition models may improve the reliability of the system. Additionally, the challenge of head positioning could be mitigated by incorporating a calibration process prior to each session or by employing an adaptive approach that adjusts according to user input during the session.

Overall, this study lays the groundwork for future investigations into the use of virtual agent-assisted cognitive training for individuals with \gls{id}, and the results underscore the need for further development to refine the system's effectiveness and address technical challenges.

\section*{Acknowledgment}
We would like to thank the employees and staff of the proWerk sheltered workshop for their commitment to the project.

\bibliographystyle{IEEEtran}
\bibliography{references}

\vspace{12pt}
\color{red}

\end{document}